\def\oiii{[O~{\sc iii}]}

\documentclass[11pt,twoside]{article}  
\usepackage{apn3conf}
\begin{document}   

\title{The Luminosity Function of PNe with different morphology}

\author{Magrini L.\altaffilmark{1}, Corradi R. L. M.\altaffilmark{2}, Leisy P.\altaffilmark{2,3}, 
Scatarzi A.\altaffilmark{1}, \\Morbidelli L.\altaffilmark{4}, Perinotto M\altaffilmark{1}}

\affil{Dipartimento di Astronomia e Scienza dello Spazio, Universit\'a di     
Firenze, L.go E. Fermi 2, 50125 Firenze, Italy}     
\affil{Isaac Newton Group of Telescopes, Apartado de Correos 321, 38700 Santa       
Cruz de La Palma, Canarias, Spain}
\affil{Instituto de Astrof\'{\i}sica de Canarias, c. V\'{\i}a L\'actea s/n,      
38200, La Laguna, Tenerife, Canarias, Spain}         
\affil{IRA/CNR L.go Fermi, 5 50125 Firenze, Italy}

\contact{Laura Magrini }
\email{laura@arcetri.astro.it }
\paindex{Magrini, L. }
\aindex{Corradi, R. L. M.}
\aindex{Leisy, P.}
\aindex{Scatarzi, A.}
\aindex{Morbidelli, L.}
\aindex{Perinotto, M.}
\authormark{Magrini, Corradi, Leisy, Scatarzi, Morbidelli \& Perinotto} 

\begin{abstract}      
We have analyzed the behaviour of various parameters of PNe in the
Magellanic Clouds (MCs) and the Galaxy as a function of their
morphology.  The luminosity function of different morphological types
has been built, finding that elliptical and round PNe dominate the
bright cutoff both in the MCs and in the Galaxy.  The dependence of
the ${\rm [OIII]}$ absolute magnitude on chemical abundances has been
investigated.
\end{abstract}

\section{The samples}
The MCs sample (51 objects) has been selected choosing PNe whose
morphology was studied with the {\sl HST} by Stanghellini et
al. (1999, 2002a) and Shaw et al. (2001). Their chemical abundances
and relative fluxes have been obtained from the work of Leisy \&
Dennefeld (2003), and the absolute fluxes from Jacoby et
al. (1990). \\ We have analyzed two Galactic samples as well. We have
obtained their morphological classification from Corradi \& Schwarz
(1995) and Stanghellini et al. (2002b).  The first sample is composed
by PNe whose data are available in the ESO-Strasbourg Catalogue (Acker
et al. 1992), while the second one consists of  objects in common 
between PNe whose chemical abundances have
been re-determined by Perinotto et al. (2003) and PNe whose distances have
been recently obtained by Phillips (2002). We consider the distances
of the second sample to be more uniform and  accurate.

\section{The Planetary Nebulae Luminosity Function}

We have investigated the dependence of the PNe luminosity function on
their morphology in order to study which kind of PN has
the highest probability to be observed in a far galaxy where only the
brightest part of the luminosity function can be detected.
\begin{figure}
\epsscale{.6}
\plotone{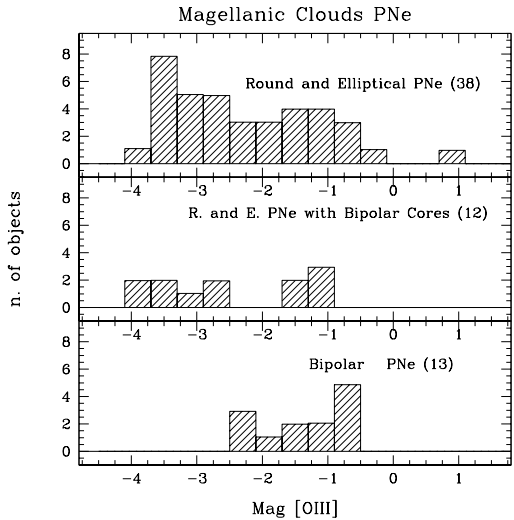}
\caption{The  \oiii\ 5007\AA\ luminosity function of MCs PNe whose morphology has been 
studied} 
\end{figure}
In Fig.~1 and Fig.~2 we present the \oiii\ 5007\AA\ luminosity
function of the MCs and Galactic PNe respectively, built distinguishing among different
morphological types. \\ From these figures, we have noted that the
PNLF bright cutoff is dominated by elliptical and round PNe both in
the the MCs and in the Milky Way.  Elliptical PNe with bipolar cores
(a class defined by Stanghellini et al. 1999, but for which it is not clear whether they
really constitute an independent physical class of PNe) appear to have 
the same maximum \oiii\ luminosity  as normal ellipticals and rounds.
\begin{figure}
\epsscale{.90}
\plottwo{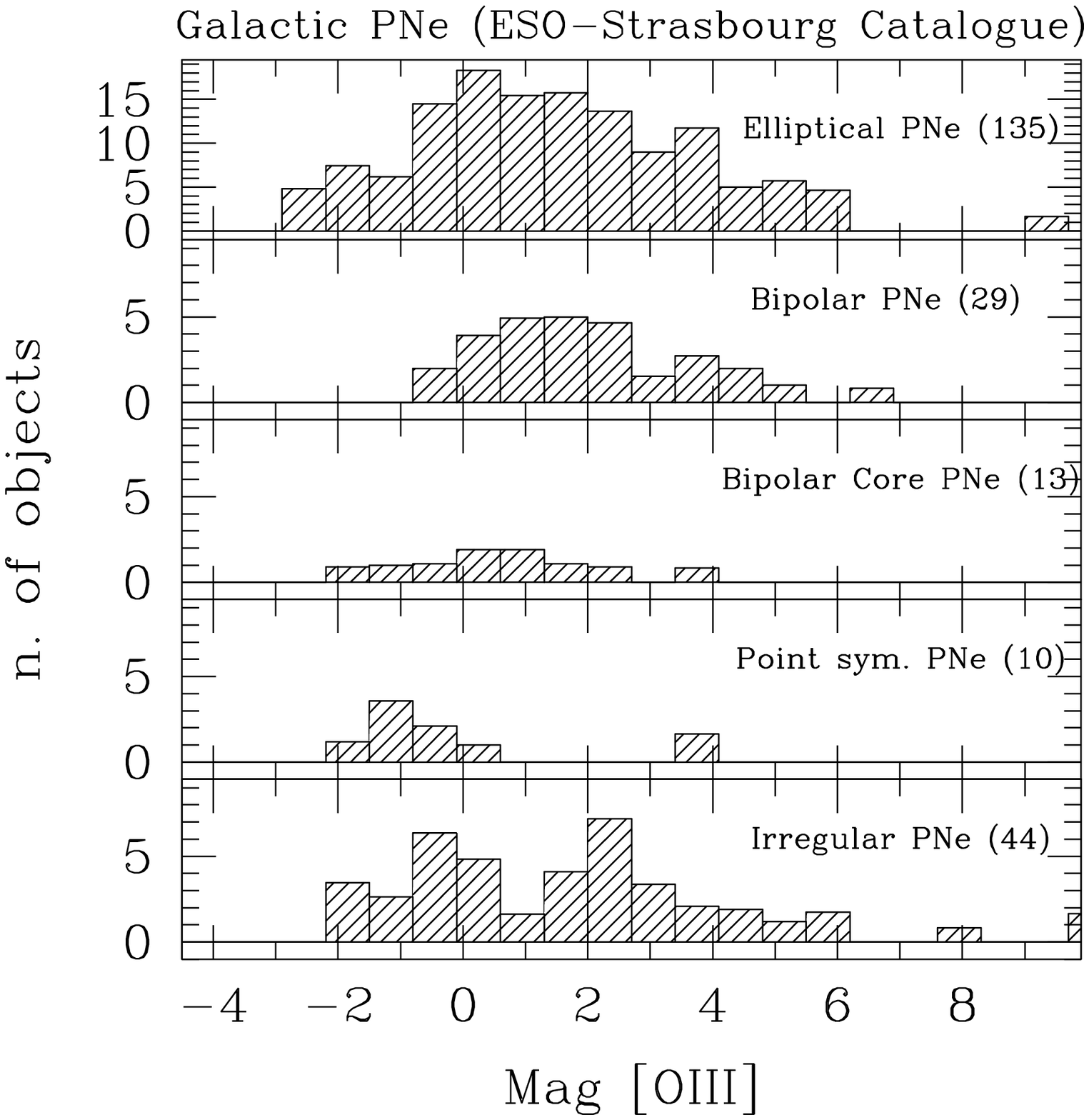}{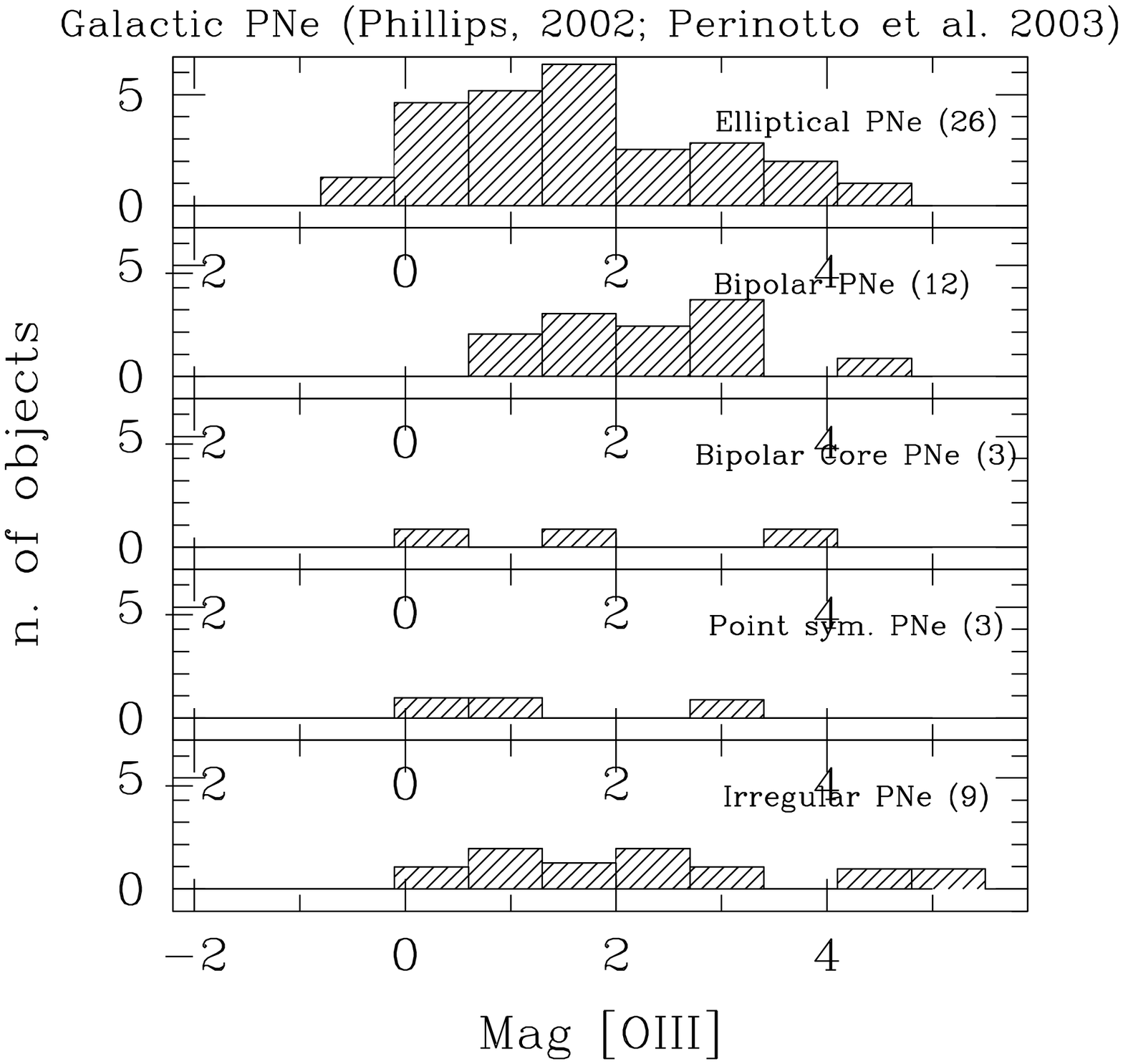}
\caption{The \oiii\ 5007\AA\ luminosity function of a sample of Galactic PNe. 
On the left (a) the data from the ESO-Strasbourg Catalogue (Acker et
al. 1992).  On the right (b) the PNe with the best chemical abundances
(Perinotto et al. 2003) and the most recent distance determinations
(Phillips 2002).}
\end{figure}
We have noted that in the Galaxy and in the MCs, bipolar PNe are generally
less bright (in the \oiii5007\AA\ line) then other morphological
types.  This behaviour can be interpreted considering that bipolar PNe
are generally associated with more massive progenitors (cf. Corradi \&
Schwarz 1995).  In fact, according to stellar evolution
models, after ejection of the envelope at the tip of AGB the stellar
core evolves to  higher temperatures and subsequently turns down in the H-R
diagram moving more slowly toward the white dwarf area.
Higher core masses, which come from higher initial stellar 
masses,  have  larger  initial luminosities and  reach greater temperatures,
but also have  faster evolutions.  Consequently  they  are expected 
to be found on average at higher temperatures but not at higher luminosities 
than lower mass objects (cf.  M\'endez et al. 1993).

\section{Chemical Abundances}
The determination of chemical abundances in a far galaxy from the
information contained in spectra of PNe are generally obtained
observing the brightest objects.  We have examined the chemical abundances
of PNe with different morphological types {\sl vs} mag \oiii\ in LMC
(the sample of SMC PNe was too limited) and in the Galaxy to test for
possible biases in measuring abundances from the brightest PNe (see
Fig.~3a-b).  In fact, as said in the previous section, the brightest
PNe usually belong to elliptical or round (with or without internal
bipolar structures) morphological classes. Note also that, in any
classes of PNe, the chemical abundances of the element which were not
altered during their evolution, i.e. Ne, Ar, S, represent the
chemistry of the interstellar medium of the host galaxy at various
times.
\begin{figure}
\epsscale{1.}
\plottwo{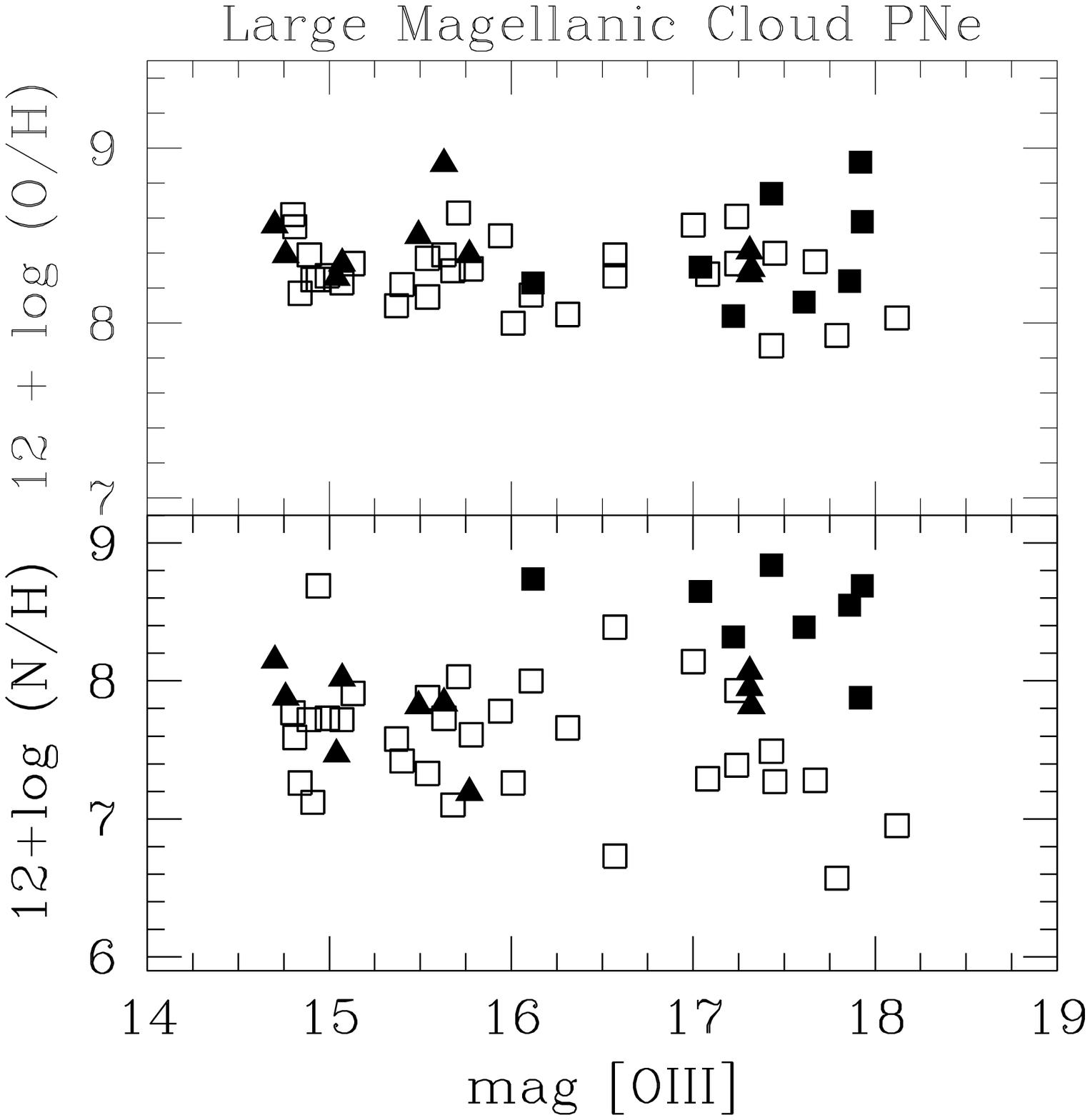}{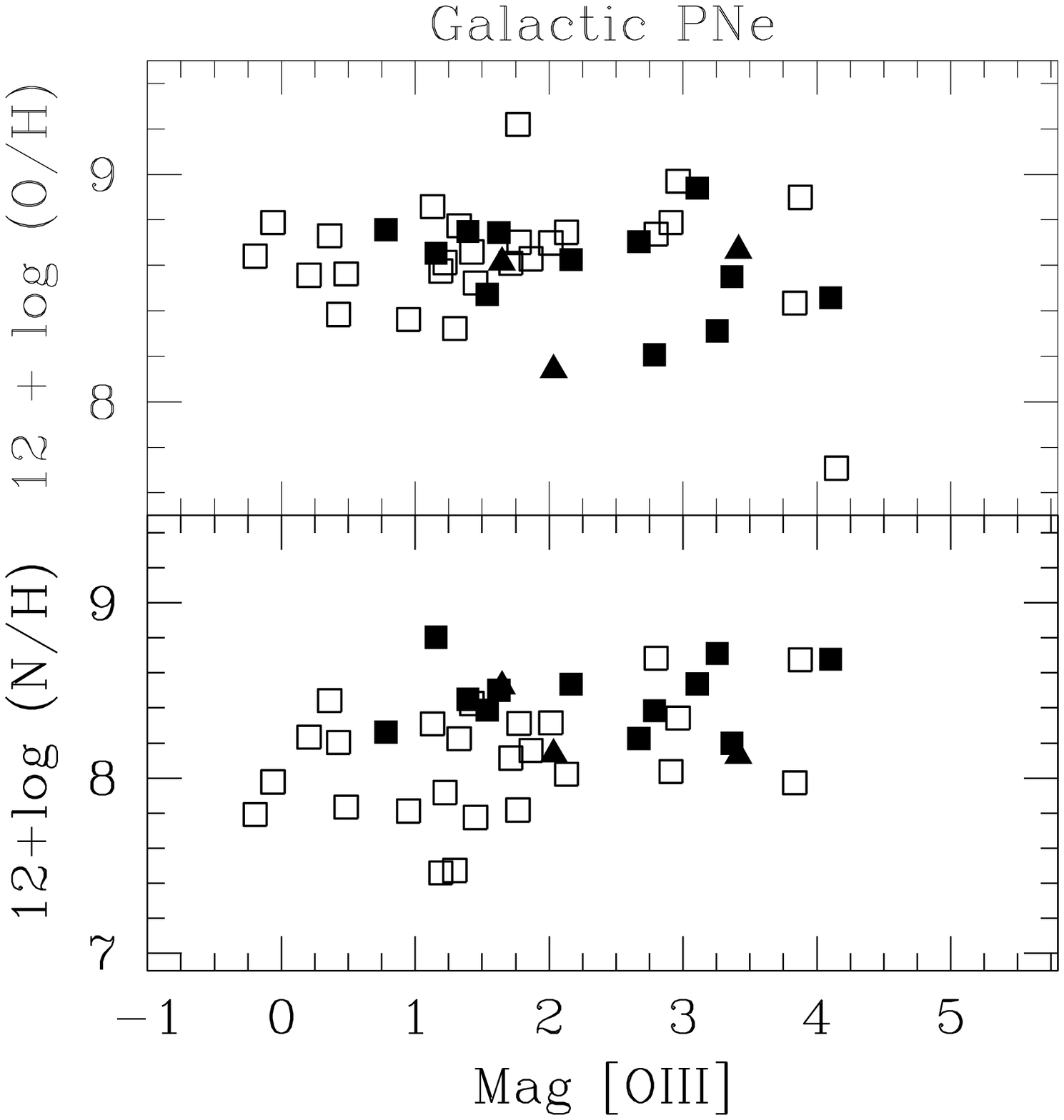}
\caption  {\oiii\  magnitude {\sl vs} chemical abundances of a sample of LMC (left, Fig.~3a) and 
of Galactic PNe (right, Fig.~3b). 
Filled squares are bipolar PNe,  triangles are elliptical or round PNe with a bipolar core 
 and empty squares are elliptical or round PNe.}
\end{figure}
In Fig.~3a-b we looked at the chemical behaviour of O and N
in the LMC and in the
Galaxy.  In the LMC PNe we notice that the behaviour of these chemical
abundances {\sl vs} \oiii\ luminosity is quite similar for elliptical
and round PNe with and without bipolar core (triangles and empty
squares). 
The bipolar morphological class (filled squares) is
represented almost by type I PNe (cf. Perinotto \& Corradi 1998),
which are overabundant in nitrogen and helium with respect to
ellipticals.  An increasing trend of the S, Ar, Ne, and O {\sl
vs} \oiii\ magnitude is seen (see Fig.3a for O) in the bipolar sample. 
This might be due  to the
shorter life of the younger and more massive progenitors which reach
quickly lower luminosities and come from a more enriched interstellar
medium.

In the sample of Galactic PNe (with chemical abundances from Perinotto
et al. 2003 and distances from Phillips 2002, Fig.~3b for O and N) we also note
that bipolar PNe are on average overabundant in N and He.  For the
considered sample of Galactic PNe, no evident trend of chemical
abundance {\sl vs} \oiii\ luminosity is seen.  Here, however, the
uncertainties on distances which affect  the \oiii\ luminosity are
larger than those in the LMC sample.  In addition, Galactic gradients
of the chemical abundances spread the data over a large range.

In both cases, LMC and Galaxy, we conclude that the chemical
abundances derived from the brightest PNe are 
representative of the total PNe population.  In fact, no strong
changes of chemical abundances with \oiii\ luminosity are seen, except
for the well-known overabundance of N and He in bipolar PNe.


\end{document}